# Time-Varying Directed Interactions in Functional Brain Networks: Modeling and Validation

Nan Xu[1,2], Xiaodi Zhang[2], Wen-Ju Pan[2], Jeremy L. Smith[3], Eric H. Schumacher[4], Jason W. Allen[3,5], Vince D. Calhoun[6,2], Shella D. Keilholz[2]

[1]Fischell Department of Bioengineering, Department of Electrical and Computer Engineering, Neuroscience and Cognitive Science Program, University of Maryland, College Park, MD, United States
[2]Wallace H. Coulter Department of Biomedical Engineering, Georgia Institute of Technology and Emory University, Atlanta, GA, United States
[3]Department of Radiology and Imaging Sciences, Emory University School of Medicine, Atlanta, GA, United States
[4]School of Psychology, Georgia Institute of Technology, Atlanta, GA, United States
[5]Department of Radiology and Imaging Sciences, Indiana University School of Medicine, Indianapolis, IN, United States
[6]Tri-institutional Center for Translational Research in Neuroimaging and Data Science, Georgia State University and Georgia Institute of Technology and Emory University, Atlanta, GA, United States

*Abstract*

Understanding the dynamic nature of brain connectivity is critical for elucidating neural processing, behavior, and brain disorders. Traditional approaches such as sliding-window correlation (SWC) characterize time-varying undirected associations but do not resolve directional interactions, limiting inference about time-resolved information flow in brain networks. We introduce sliding-window prediction correlation (SWpC), which embeds a directional linear time-invariant (LTI) model within each sliding window to estimate time-varying directed functional connectivity (FC). SWpC yields two complementary descriptors of directed interactions: a strength measure (prediction correlation) and a duration measure (window-wise duration of information transfer). Using concurrent local field potential (LFP) and fMRI BOLD recordings from rat somatosensory cortices, we demonstrate stable directionality estimates in both LFP band-limited power and BOLD. Using Human Connectome Project (HCP) motor task fMRI, SWpC detects significant task-evoked changes in directed FC strength and duration and shows higher sensitivity than SWC for identifying task-evoked connectivity differences. Finally, in post-concussion vestibular dysfunction (PCVD), SWpC reveals reproducible vestibular-multisensory brain-state shifts and improves healthy-control vs subacute patient (HC-ST) discrimination using state-derived features. Together, these results show that SWpC provides biologically interpretable, time-resolved directed connectivity patterns across multimodal validation and clinical application settings, supporting both basic and translational neuroscience.

# 1. Introduction

The human brain is a highly dynamic system, where functional connectivity (FC) evolves over time (Allen et al., 2012; Hutchison et al., 2013) and can exhibit asymmetric, directed interactions across brain regions (Friston, 2011; Mitra et al., 2015; Xu et al., 2021). Capturing these temporal and directional dynamics is critical for understanding neural processing, behavior, and the mechanisms underlying brain disorders. Traditional FC methods, such as sliding window correlation (SWC), have been invaluable for investigating temporal variability in functional magnetic resonance imaging (fMRI) data (Allen et al., 2012; Shakil et al., 2016). By segmenting time series data into overlapping windows, SWC enables researchers to explore how connectivity fluctuates over time, providing critical insights into brain network dynamics associated with cognition and disease. However, SWC is inherently limited to correlational analyses and lacks the ability to infer directionality in information flow. These limitations constrain its utility in characterizing hierarchical neural interactions, which are essential for understanding the mechanistic basis of both normal brain function and clinical conditions, such as post-concussion vestibular dysfunction (PCVD) (Smith et al., 2021).

In neuroimaging literature, methods for studying causality and correlation have long been viewed as distinct and divided, reflecting the differing goals and assumptions underlying each approach (Aedo-Jury et al., 2020; Friston, 2011; Friston et al., 2003; Keilholz et al., 2013; Liang et al., 2015; Park et al., 2018; van den Heuvel and Hulshoff Pol, 2010). Correlational approaches, such as SWC, focus on statistical dependencies without inferring directional influence, whereas causal models aim to uncover mechanistic interactions driving these dependencies. This conceptual separation arises from different goals: correlational methods measure symmetric statistical dependence without directionality, whereas causal models focus on asymmetric influence by distinguishing cause from effect (Peters et al., 2017). Recent work has advanced time-varying effective-connectivity modeling for EEG/MEG, often demonstrated in relatively low- to moderate-dimensional channel/source settings, e.g., (Medrano et al., n.d.). Complementary developments are still needed to enable scalable, whole-brain, time-resolved modeling for resting-state fMRI, where hemodynamic filtering and high dimensionality impose distinct constraints. This practical gap has constrained efforts to integrate correlational and directed perspectives within a unified, time-resolved framework.

To bridge this gap, we propose sliding window prediction correlation (SWpC), a computational framework that embeds a validated causal linear time-invariant (LTI) model (Xu et al., 2017) within each sliding window. For each window and each direction $x \to y$, SWpC predicts $y$ from $x$ via an LTI impulse response, then quantifies directed strength using prediction correlation: the correlation between predicted and observed $y$ in that window. In parallel, SWpC estimates a duration of information transfer as the window-specific impulse-response support (chosen by model selection), providing a time-resolved measure of how temporally extended the influence of $x$ on $y$ is. Importantly, strength and duration need not covary: interactions can be brief but strong, or weaker yet temporally sustained. Together, these complementary measures allow SWpC to characterize time-varying directional interactions while retaining the practical sliding-window workflow used in dynamic FC studies.

Validation of directionality estimates from BOLD fMRI is particularly critical due to the variability in hemodynamic responses across brain regions. While prior studies have validated direction estimation using simulated data (Smith et al., 2011; Xu et al., 2017), the inherent limitations of simulations, including oversimplified assumptions, underscore the need for validation with biologically relevant data. To address this challenge, we leverage a diverse range of multimodal neuroimaging datasets, including concurrent LFP-BOLD recordings in rodents, human task-based fMRI, and patient fMRI data, which serve as robust benchmarks for testing the reliability and sensitivity of the proposed model.

Using high-content LFP-BOLD datasets, SWpC demonstrated its ability to reliably estimate directed interactions at both neuronal and BOLD levels, extending previous SWC results by adding directionality in strength and duration. Specifically, SWpC strength and duration were consistently estimated in both BLP and BOLD signals from S1L and S1R, exhibiting symmetries between the two regions, as their directional asymmetry consistently remained below scan variability. Additionally, SWpC strength reflects LFP-BOLD correlations akin to those uncovered by SWC, with high BLP-BOLD correlates observed within the $\theta$, low $\beta$, and upper bands (Garth John Thompson et al., 2013a), while SWpC duration exhibited only moderate LFP-BOLD correlation. These findings underscore SWpC's ability to capture both dynamic and directional connectivity with biological plausibility.

SWpC's sensitivity was further validated using human motor task fMRI data from the Human Connectome Project (HCP). The method identified significant task-driven changes in directional FC strength and duration, uncovering asymmetries between task and rest conditions with large effect sizes. Compared to SWC, SWpC showed greater sensitivity in detecting directional interactions, underscoring its reliability in dynamic FC analysis. Finally, we assessed the potential clinical utility of SWpC by testing whether brain-state features derived from resting-state fMRI can reliably distinguish subacute PCVD patients from healthy controls. SWpC identified reproducible vestibular–multisensory brain states and achieved stronger HC–ST discrimination with strength-based features than SWC, whereas duration-based features offered limited separability. By capturing these task-specific and clinical insights, SWpC highlights its potential to advance basic neuroscience and clinical research.

This hybrid approach bridges the gap between causality and correlational analysis, offering a unified framework for analyzing time-varying directional connectivity in brain networks. Its successful application across diverse datasets demonstrates its robustness, sensitivity, and clinical utility, advancing our understanding of neural dynamics in both health and disease.

## 2. Results

### 2.1. Sliding-window prediction correlation (SWpC) model

The model integrates the sliding-window technique with a validated static causal model of prediction correlation (Nan Xu et al., 2017; Xu et al., 2021) to analyze time-varying causal interactions in brain networks. Conceptually, in the causal and time-varying functional brain networks (FBNs), every directed functional connectivity (FC) interaction (e.g., x→y and y→x) is modeled as an independent communication channel within a causal, time-varying system.

To ensure computational scalability, the sliding-window correlation (SWC) approach approximates the time-varying system by dividing the time course into short, overlapping windows, each treated as a time-invariant model. Within these windows, a static causal system (e.g., a linear time-invariant (LTI) model as outlined in (Nan Xu et al., 2017; Xu et al., 2021)) is embedded for analysis. For a specific window $[\tau, \tau + L]$ (starting at time $\tau$ with length L), the directed FC can be computed in the following two steps:

1) Prediction: The windowed segment of y (denoted by $y_\tau$) can be predicted by an LTI model, $\hat{y}_\tau[n] = x_\tau[n] * h_\tau[n] = \sum_{m=0}^{D_{y|x}[\tau]-1} h_\tau[m] x_\tau[n-m]$, for $n \in [\tau, \tau+L]$, where $x_\tau$ is an x segment within the current and the previous windows, and $h_\tau[n]$ is the window-specific impulse response with duration $D_{y|x}[\tau]$

2) Correlation: The strength of directed FC is calculated as the correlation between the predicted and observed segments: $\rho_{y|x}[\tau] = corr(\hat{y}_\tau, y_\tau)$.

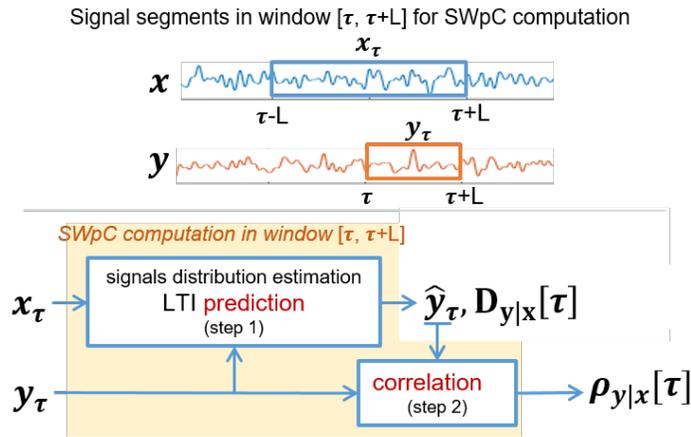

**Fig. 1 SWpC computation in each window.**

Notably, in step (1), the input signal x in the previous window is also included for capturing the causal influences (see Fig. 1). Due to the previous success of prediction correlation in (Nan Xu et al., 2017; Xu et al., 2021), the LTI system can be reliably predicted for each window. The model predicts time-varying directed FC, with both strength ($\rho_{y|x}[\tau]$) and duration ($D_{y|x}[\tau]$) depending on the window's starting point τ, enabling directional distinctions ($\rho_{y|x}[\tau] \neq \rho_{x|y}[\tau]$ and $D_{y|x}[\tau] \neq D_{x|y}[\tau]$ for any τ's). This dynamic method, namely sliding window prediction correlation (SWpC), extends SWC by incorporating causal influences and distinguishing between forward and backward connections, offering a more nuanced analysis of brain connectivity.

## 2.2. SWpC reliably predicts directional strength and duration from both LFP and BOLD signals.

The strength and duration of SWpC were reliably estimated from both BLP and BOLD signals recorded in S1L and S1R of rats (Fig. 2a), highlighting symmetrical connectivity between the two regions. Notably, the 75% quantile of directional asymmetry remained lower than the scan-to-scan variability (Fig. 2b). Figure 2B shows the distribution of mean asymmetry values across scans for each signal type, with the scan variability overlaid as a reference line. P-values derived from the Wilcoxon signed-rank test were all well below 0.05, confirming that the observed asymmetries in both strength and duration measures were significantly smaller than the scan variability, thus attesting to the stability of SWpC estimates. Furthermore, SWpC strength captured LFP-BOLD relationships similar to those revealed by SWC, showing strong BLP-BOLD correlations within the $\theta$, low $\beta$, and higher frequency bands (with mean correlations >0.1, Fig. 2c). This finding is consistent with prior results obtained from concurrently recorded LFP-BOLD data in rats under isoflurane anesthesia (Garth John Thompson et al., 2013a). In contrast, SWpC duration displayed comparatively modest LFP-BOLD correlations (with mean correlations <0.1, Fig. 2c).

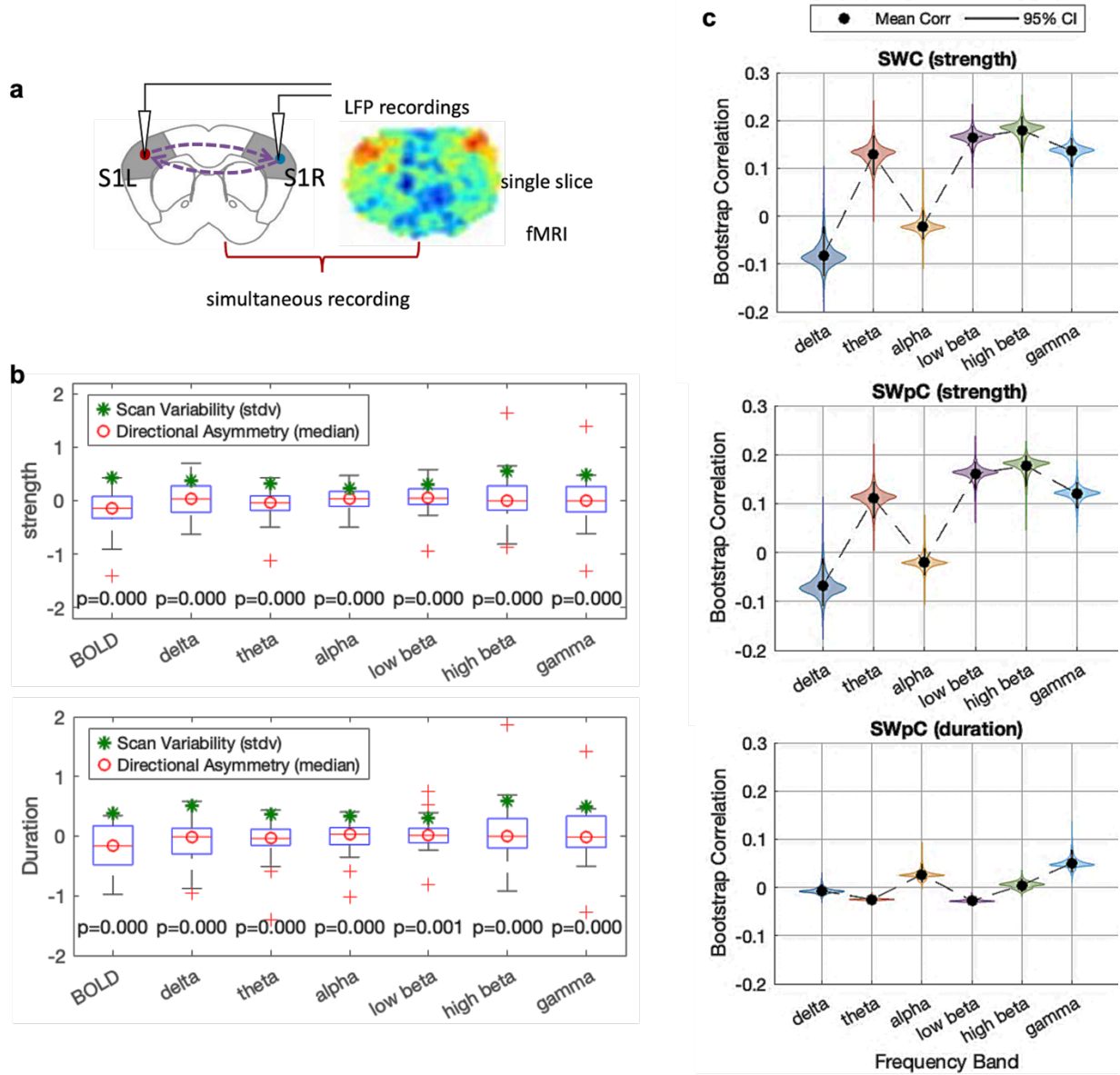

**Fig. 2 Reliability of the SWpC Measure in Simultaneously Recorded LFP-BOLD Data**. **a** Schematic of simultaneous LFP and BOLD signal recordings from bilateral somatosensory regions (S1L and S1R). **b** Boxplots comparing directional asymmetry in connectivity (S1L→S1R vs. S1L←S1R) with scan-to-scan variability in SWpC strength (left) and SWpC duration (right) estimates for each signal type. The red line and red circle indicate the median directional asymmetry within each boxplot, while the green star denotes the corresponding scan-to-scan variability. **c** Bootstrap correlation analysis between BOLD results and band-limited power (BLP) results in six frequency bands for three metrics: SWC strength (left), SWpC strength (middle), and SWpC duration (right). For SWpC result, S1L→S1R is used as an illustrative example.

## 2.3. SWpC sensitively predicts directional functional connectivity from motion-evoked human fMRI data

As described in the Method section, the data were obtained from 45 subjects in the HCP test-retest dataset, where participants were cued to move their left or right foot, hand, or tongue during two repeated runs. Following the 3-level Feat analysis, we identified 49 ROIs (listed in Fig. S3) that align with previously reported motor task activation maps and connectome fingerprints covering the s (Barch et al., 2013; Tripathi et al., 2024). Specifically, foot activations are located on the superior midline, hand activations ventral to the foot, and tongue activations ventral to the hand (Fig. S2–4). In the cerebellum, ipsilateral hand/foot and bilateral tongue representations contrast with the contralateral patterns observed in the motor cortex.

As shown in Fig 3a and Fig S5, the SWpC results demonstrate that motion enhances directional information flow compared to the corresponding rest trials, which appear to exhibit more symmetrical connectivity patterns as shown in SWC results. This enhancement is evident in both strength and duration, with SWpC uniquely capturing temporal aspects of directed connectivity that are not addressed by SWC. Across all motion types, directional asymmetry in strength and duration during task performance was significantly greater than during rest, with Cohen's $d$ values exceeding 0.8 (see mean asymmetry and Cohen's $d$ values in Table S1), indicating large effect sizes and substantial differences between the two conditions. The violin plots in Fig. 3b further emphasize these findings, showing consistently increased asymmetry in directed FC during tasks relative to rest, in both strength and duration, across all motions.

Directed FC durations predicted by SWpC revealed significant differences between task and rest trials (FDR-corrected $p < 0.01$). Specifically, directed FC with significantly longer durations during tasks performances, coupled with low individual variability (CV < 36%), identified hub-like structures associated with hand and foot movements (Fig. 3c). Large ROIs in somatomotor areas responding to these movements exhibited increased long-range information transfer from other ROIs, as indicated by red and green arrows highlighting significant differences for left and right motions, respectively. Additionally, the averaged time series for each task, across all trials and subjects, revealed that these ROIs exhibited delayed and higher hemodynamic peaks compared to other ROIs (Fig. 3a, bottom row, and Fig. S5c). This temporal delay may account for the observed increase in the duration of information transfer.

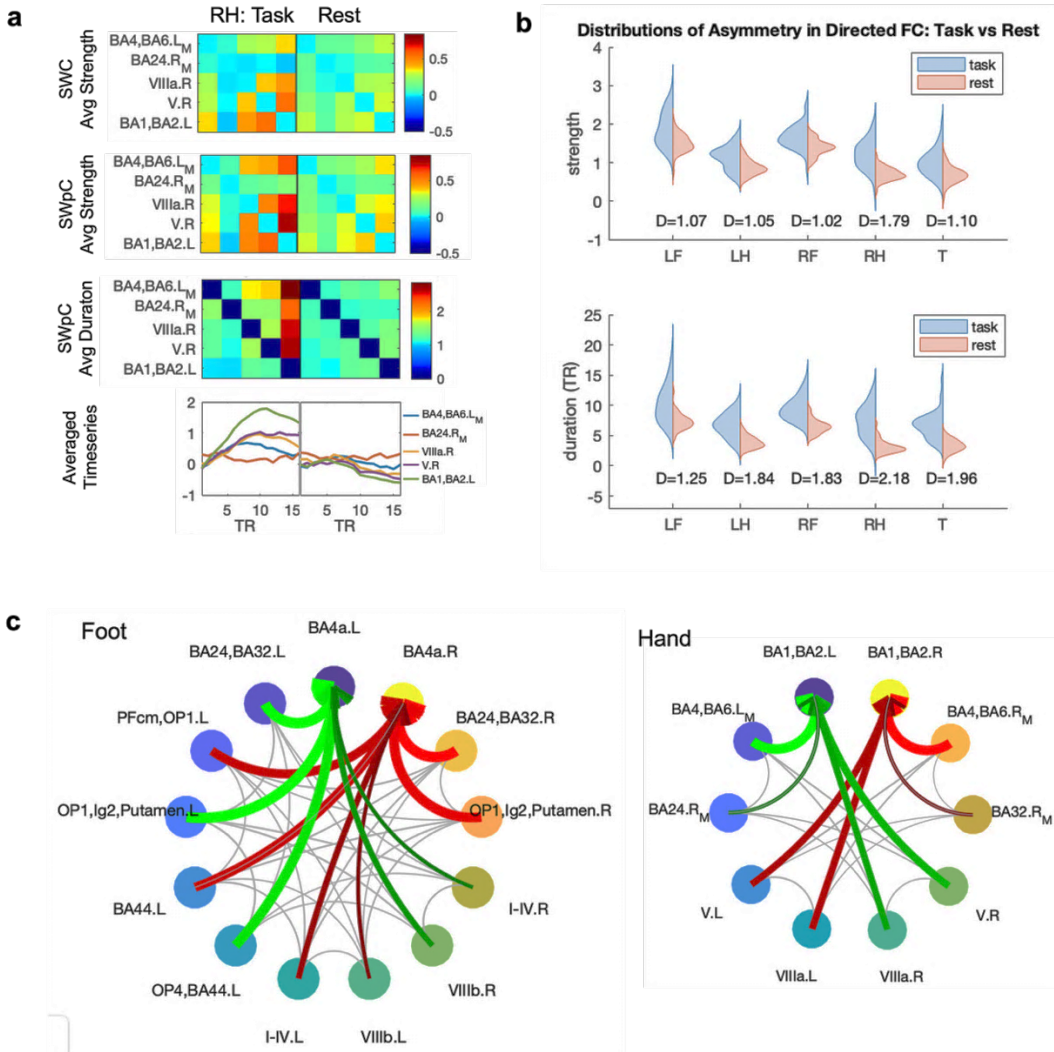

**Fig. 3 Asymmetry in directed FC and significant long durations between task vs. rest. a** Example of group-averaged SWpC results for right-hand (RH) task, illustrating enhanced directional information flow compared to rest trials. All task-specific results are presented in Fig. S5. The top row presents the averaged FC strength from sliding window correlation (SWC). The second and third rows display the averaged directed FC strength and duration estimated by sliding window prediction correlation (SWpC), respectively. For simplicity, only larger right-hand ROIs (≥25 voxels) are included. The bottom row shows the averaged time series across all task (or all rest) trials for the five right-hand regions of interest (ROIs). **b** Violin plots summarizing directional asymmetry in strength and duration for all tasks, showing consistently increased asymmetry during task performance relative to rest, with large effect sizes (Cohen's d > 0.8). The corresponding Cohen's *d* value is listed under each pair of violin plots, indicating the magnitude of the difference in means of task and rest distributions relative to their pooled standard deviation (with values ≥0.8 generally considered large effects). **c** Connectograms of directed FC with significant differences in duration between task and rest for hand (left) and foot (right) motions. Red (green) arrows indicate task vs. rest differences for left (right) motion with low individual variability (CV < 36%). Gray connections represent all predicted (directed) connections based on SWpC. For clarity, the connectograms include 13 ROIs for foot and 10 ROIs for hand motions, excluding smaller ROIs (fewer than 100 voxels for foot motions and fewer than 30 voxels for hand motions). Remaining ROIs demonstrate approximate symmetry for left/right foot and hand tasks. Under these thresholds, tongue-evoked information flow among the 4 larger ROIs does not form hubs in the motor cortex as shown in Fig. S6.

Significant differences between the motion task and rest trials were also observed in the SWpC-predicted directed FC strengths. Figure 4 highlights the significantly enhanced directed FC strength evoked by each task, characterized by low individual variability (FDR-corrected $p < 0.01$, CV < 30%), contrasting with the symmetrical patterns estimated by SWC. For each of the hand and foot tasks, SWpC detected significantly evoked causal interactions between the cerebellum and the contralateral somatomotor regions, aligning with the SWC undirected FC results. Examples include VIIIb.L → BA4a.R for left foot motion, VIIIb.R → BA4a.L for right foot motion, VIIIa.L → BA1, BA2.R for left-hand motion, and VIIIa.R → BA1, BA2.L for right-hand motion. Additionally, SWpC demonstrated its sensitivity in detecting information flow from the OP1, Lg2, and putamen to BA4a for both left and right foot movements. Consistent with current understanding of sensorimotor control and neural connectivity, these connections were more lateral for hand than foot movements and most lateral for the tongue movement task (Fieblinger, 2021; Rizzolatti and Luppino, 2001).

The observed information flow from the medial extensional region of BA4/BA6 to BA1/BA2 during hand movements aligns with established evidence of the motor-sensory integration necessary for precise motor execution. This flow reflects the hierarchical motor control process, where motor planning and execution (BA4/BA6) are tightly coupled with sensory feedback processing (BA1/BA2) (Todorov, 2004). The ability of SWpC to detect such task-evoked directed functional connectivity underscores its sensitivity and reliability in identifying biologically meaningful information flow, as highlighted in effective connectivity studies of hierarchical motor control (Fieblinger, 2021).

Moreover, SWpC sensitively detected bilateral information flows from the cerebellum to contralateral tongue-specific regions in the somatomotor cortex (VI.L → BA3b, OP4.R; VI.R → BA3b, OP4.L), whereas SWC identified only one pair under the same p-value and CV thresholds. These findings are consistent with the well-established role of the cerebellum in coordinating fine motor task (Manto et al., 2011), including precise bilateral movements required for tongue control (Sasegbon and Hamdy, 2023). By capturing these subtle dynamics, SWpC demonstrates superior sensitivity, with over 10% greater detection of task-evoked connections compared to SWC (Fig. 4c).

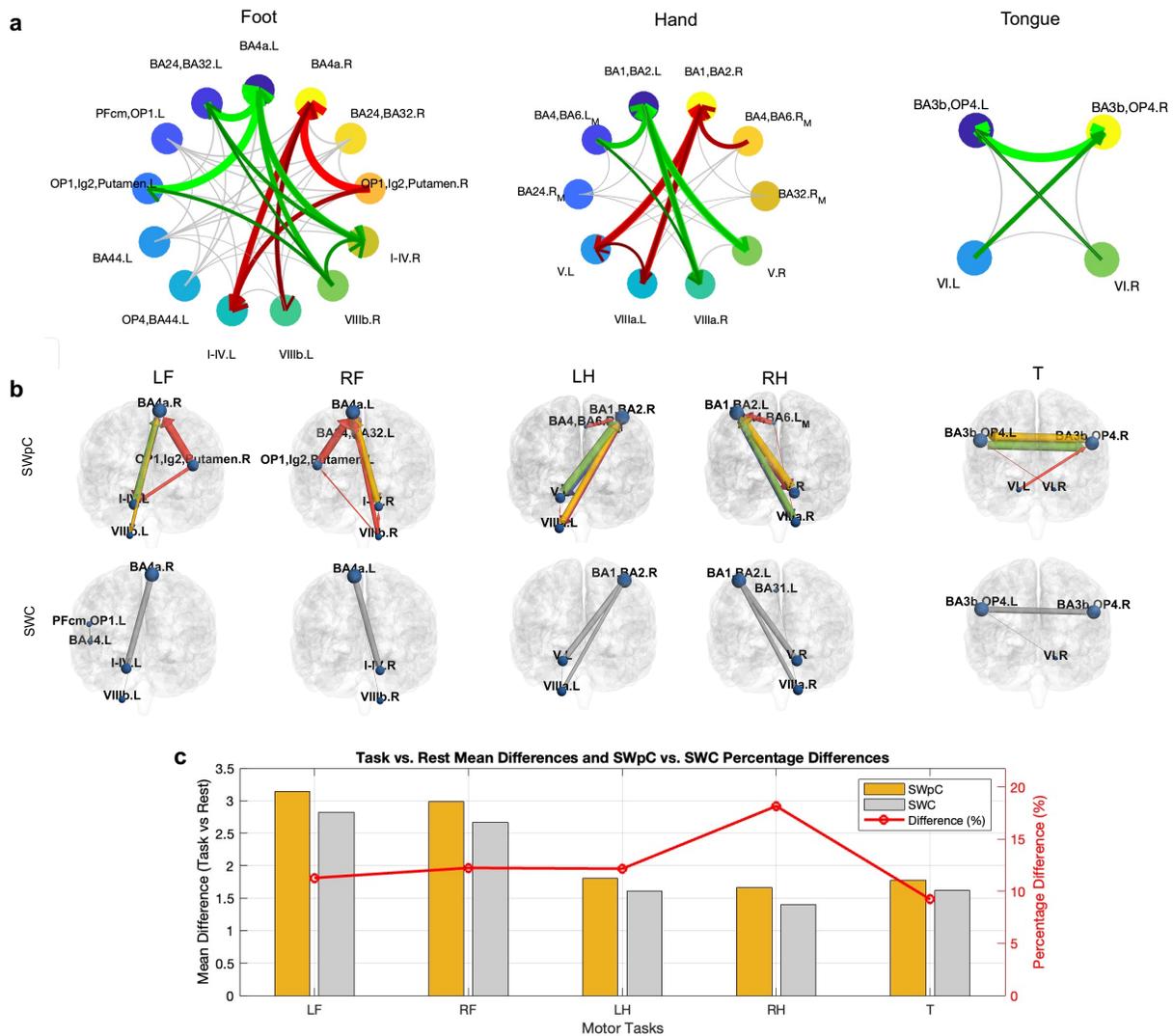

**Fig. 4 Directed FC with significant strength differences between task and rest for each motion.**
**a** Connectograms showing directed FC strengths significantly evoked by each motion (FDR-corrected p < 0.01). For hand and foot motions, red (green) arrows depict significant directed FC strengths for left (right) motions with low individual variability (CV < 30%). For tongue motions, green arrows represent significant evoked directed FC strengths with low variability. Gray connections denote all predicted directed connections from SWpC. To enhance clarity, connectograms include 13 foot ROIs, 10 hand ROIs, and 4 tongue ROIs, excluding smaller ROIs (fewer than 100 voxels for foot motions and fewer than 30 voxels for hand and tongue motions). The remaining ROIs exhibit approximate symmetry for left/right foot and hand tasks. **b** Brain regions and directed functional connections with significantly task-evoked strength predicted by SWpC are visualized and contrasted with SWC results. For both methods, connections with significant task-rest strength differences (FDR-corrected p < 0.01) and low variability (CV < 30%) are shown. For SWpC, these connections correspond to the red and green arrows in the connectogram in **a**. Arrow thickness represents connectivity strength, while the arrow color (cool to warm) indicates connection duration from short to long in SWpC results. **c** A bar plot comparing the mean strength differences between task and rest estimates for SWpC and SWC. The percentage difference, calculated as (SWpC−SWC)/SWC×100%, is shown as a red line.

Furthermore, temporal variations in SWpC strength were strongly correlated with SWC strength variations across all motion tasks (Pearson's r = 0.85, p < 0.001), as shown in the scatter plot (Fig. S7). This strong correlation demonstrates consistency between the two methods in capturing FC strength temporal dynamics. The correlation between the group-averaged mean FC strength (calculated as the average of all connections within the group-averaged FC matrix) across the four trials (Fig. S8) and the group-averaged reaction time (RT) for each task is presented in Table 1. Despite the small sample size (n = 4) and the limited variability in RTs for these motions (Fig. S8), SWpC consistently demonstrated a reverse correlation with RT across all motion tasks, aligning with expectations. Notably, SWpC demonstrated greater sensitivity compared to SWC, particularly in tasks such as LH and T, where SWC yielded marginal positive correlation with RT. These findings underscore the enhanced capability of SWpC in detecting task-specific connectivity changes while maintaining agreement with traditional methods.

|  | LF | RF | LH | RF | T |
|---|---|---|---|---|---|
| SWpC | -0.64 (0.36) | -0.96 (0.04) | -0.36 (0.64) | -0.82 (0.18) | -0.09 (0.91) |
| SWC | -0.86 (0.14) | -0.96 (0.04) | 0.13 (0.87) | -0.97 (0.03) | 0.001 (1.00) |

**Table 1: Correlation (p-value) between the subject-averaged FC strength and the mean reaction time for each motion type across four trials.**

## 2.4. Dynamic brain states in PCVD and classification performance

Using sliding-window connectivity features (SWC-strength, SWpC-strength, and SWpC-duration), we identified five reproducible dynamic brain states that capture distinct whole-network configurations across the vestibular–multisensory ROI set (Fig. 5a). Across both strength- and duration-based representations, the state structure showed two complementary (asymmetric) pairs of configurations with opposing patterns of within- versus between-network coupling. The remaining state played a more "baseline" role, but its expression depended on the metric: for SWpC-duration, state 1 was characterized by comparatively weaker and more diffuse connectivity, whereas for the strength-based representations, state 1 instead emphasized stronger within-network connections. Together, these shared states provide a common state space for summarizing time-varying vestibular–multisensory network organization in both healthy controls and PCVD patients.

Notably, SWpC-strength resolved six communities within the core vestibular ROI set, whereas the traditional SWC-strength yielded only five communities and grouped the vestibular node (VestibN) with A23c, IPLPF, PIC, IPL, and PFcm (also see Fig. S12 for group specific brain states and networks). Given the vestibular node's expected functional specificity, the finer community separation revealed by the SWpC-based measures appears more neurobiologically plausible and suggests that SWpC better preserves distinct vestibular-network organization (Frank and Greenlee, 2018). Fractional occupancy (FO) did not differ significantly between healthy controls (HC) and subacute PCVD (ST) for any state in SWC-strength (all $p \geq 0.12$) or SWpC-strength (all $p \geq 0.13$). In contrast, SWpC-duration showed a significant HC–ST difference in state 2 ($p = 0.03$), indicating that duration of causal interactions may capture complementary aspects of disease-related dynamics.

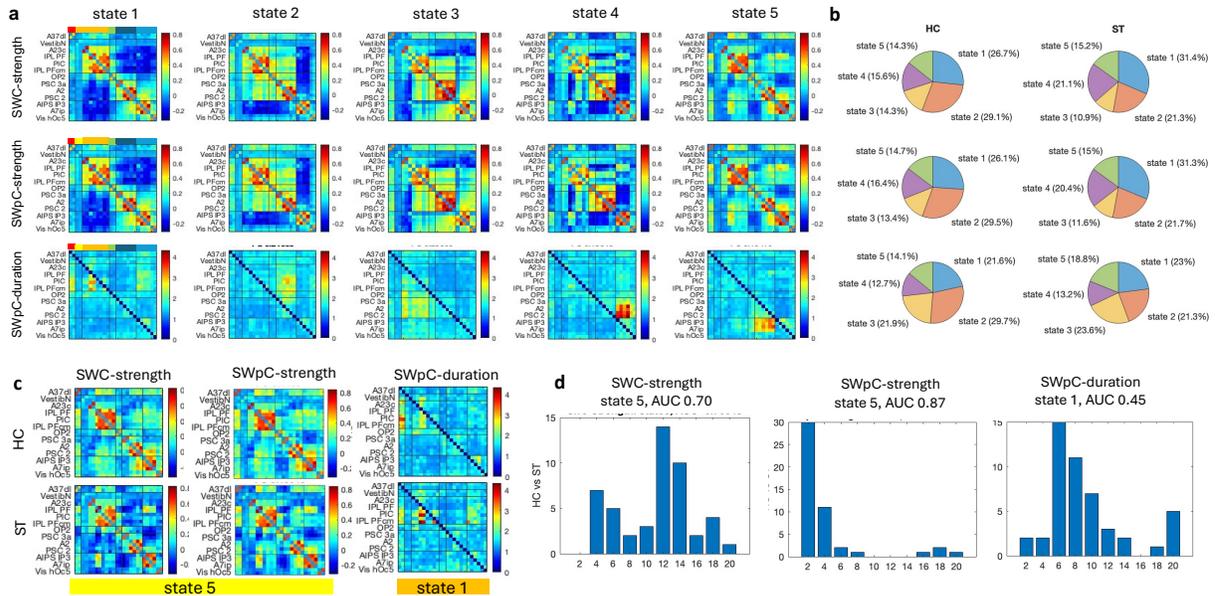

Fig. 5 **Brain-state dynamics and classification stability in PCVD. a** Five recurring states derived from the full cohort (HC, ST, CH) for SWC-strength, SWpC-strength, and SWpC-duration. **b** Fractional occupancy in HC and ST for each measure. **c** Group-averaged matrices for the states with strongest HC–ST discrimination in nested LOOCV (state 5 for SWC-strength and SWpC-strength; state 1 for SWpC-duration). **d** Threshold-selection stability across nested LOOCV: bars show how often each threshold (thr = 1–20% in 2% steps) was selected; titles report the most frequently selected state and its AUC.

In nested LOOCV, the states that most strongly discriminated subacute PCVD (ST) from healthy controls (HC) were driven primarily by connectivity-strength features rather than duration. Specifically, state 5 was most frequently selected for both SWC-strength and SWpC-strength, yielding AUC ≈ 0.70 and AUC ≈ 0.87, respectively, based on the corresponding group-averaged state matrices (Fig. 5C). Notably, SWpC-strength was more sensitive than SWC-strength, as evidenced by its higher AUC for HC–ST discrimination. This pattern is consistent with the motion-task results in the last section, where SWpC-strength likewise demonstrated greater sensitivity than SWC-strength for detecting motion-evoked connectivity changes. In contrast, for SWpC-duration, the most frequently selected state 1 exhibited near-chance performance (AUC ≈ 0.45), suggesting limited discriminative value of duration-based features in this comparison (Fig. 5D).

## 3. Discussion

In this study, we introduced sliding-window prediction correlation (SWpC), a time-resolved, direction-sensitive functional connectivity framework that embeds a causal linear time-invariant (LTI) model within each sliding window. By jointly estimating directed strength (prediction correlation) and directed duration (impulse-response support selected by model selection), SWpC extends conventional sliding-window correlation (SWC) beyond symmetric statistical dependence to capture dynamic, asymmetric interactions. Because SWpC integrates prediction correlation into a sliding-window workflow, it inherits the computational efficiency and network-level scalability of prediction-correlation (Xu et al., 2017; Xu et al., 2021), making it amenable to whole-brain analysis. Across three complementary datasets, i.e., concurrent LFP–BOLD recordings in rats, motor task fMRI in humans, and resting-state fMRI in PCVD patients, SWpC demonstrated reliability, sensitivity, and potential clinical relevance, while also revealing meaningful distinctions between strength- and duration-based descriptions of directed information flow.

### 3.1. Reliability and biological grounding from simultaneous LFP–BOLD recordings

A key challenge for directed connectivity methods is establishing biological plausibility under realistic measurement conditions, particularly for BOLD signals that are shaped by hemodynamics. Using simultaneous LFP–BOLD recordings from bilateral somatosensory cortices, we validated SWpC against a strong physiological expectation: near-symmetric interhemispheric interactions between homologous regions (Moon et al., 2025). Both SWpC strength and duration satisfied this benchmark, with directional asymmetry consistently remaining below scan-to-scan variability.

This indicates that SWpC does not spuriously impose directionality beyond what is supported by the data, and that its window-wise estimation is stable under realistic noise and nonstationarity.

Beyond stability, the multimodal recordings helped differentiate what SWpC strength and duration may capture when estimated from BOLD. SWpC strength showed BLP–BOLD correspondence consistent with prior SWC-based observations (stronger alignment in θ, low β, and higher-frequency bands), whereas SWpC duration exhibited comparatively weaker correspondence with band-limited power. A cautious interpretation is that BOLD-based duration is more susceptible to factors that shape the temporal profile of the fMRI signal such as hemodynamic smoothing, regional response timing differences, whereas BOLD-based strength more reliably reflects neural related covariation measured by LFP. This dissociation motivates treating duration as a complementary descriptor of temporal response structure rather than a direct proxy for "how long neural influence lasts."

### 3.2. Sensitivity to task-evoked directed interactions in human motor fMRI

In the HCP motor task, SWpC consistently revealed greater directional asymmetry during task than rest, indicating that task engagement produces more directionally organized interactions than those observed during the null condition. Importantly, SWpC detected task effects not only in directed strength but also in directed duration—an aspect that SWC cannot represent. Longer estimated durations formed hub-like patterns during hand and foot movements, suggesting that task engagement can expand the temporal footprint of directed interactions in addition to increasing their magnitude. Crucially, these detected information transfers were consistent with canonical sensorimotor circuit organization (e.g., cerebellar→somatomotor and effector-dependent lateralization) (Fieblinger, 2021; Manto et al., 2011; Rizzolatti and Luppino, 2001), supporting the neurobiological plausibility of SWpC-derived directionality.

At the same time, the duration findings are best understood as the BOLD timing. The longer durations likely reflect task-evoked hemodynamic timing effects: the time-averaged task-evoked BOLD responses in the task ROIs rise later and peak higher than those of other ROIs, consistent with a more temporally extended response profile (e.g., Fig. 3a). Hence, increasing SWpC duration even without implying a one-to-one mapping to prolonged neuronal influence. This interpretation aligns with the concurrent rat recordings, where duration showed weaker LFP–BOLD correspondence than strength. Taken together, these results suggest that, in task fMRI, SWpC duration is informative as a marker of temporally extended directed relationships in BOLD, potentially reflecting delayed or broadened responses in task-recruited regions, while SWpC strength may provide a more direct and stable index of directed coupling expressed in the signal.

SWpC strength additionally recovered task-evoked directed interactions that are coherent with known motor system organization, including cerebellar-to-contralateral somatomotor pathways and expected lateralization patterns across effectors. Under matched significance and variability thresholds, SWpC detected more task-evoked connections than SWC, supporting the view that prediction-based directionality can improve sensitivity while remaining neurobiologically plausible. Notably, SWpC and SWC strength variations remained strongly correlated across trials, indicating that SWpC preserves the core temporal dynamics captured by correlation-based approaches while adding interpretability in direction and temporal extent.

Finally, the exploratory association between group-averaged directed strength and reaction time consistently negative across tasks for SWpC, suggesting that stronger directed coupling may accompany faster performance. Given the small number of trial-level points (n=4), this analysis should be viewed as hypothesis-generating; nevertheless, the contrast with SWC (including marginal positive correlations in some tasks) motivates future studies with richer trial structure to test whether direction-aware measures better track behaviorally relevant network dynamics.

## 3.3. Dynamic vestibular–multisensory states in PCVD and implications for clinical utility

Applying SWpC to resting-state fMRI in PCVD patients, we identified a shared set of five reproducible vestibular–multisensory brain states across HC, subacute PCVD, and chronic PCVD (the latter included to stabilize state estimation). The state organization showed structured and interpretable configurations, including complementary asymmetric pairs that trade off within- versus between-network coupling, suggesting that directed dynamics provide a meaningful state space for summarizing vestibular–multisensory network organization.

Two findings are particularly relevant for clinical interpretation. First, SWpC-strength produced a finer modular decomposition within the vestibular ROI set than SWC-strength, separating the vestibular node into a distinct community rather than merging it with multiple parietal/insular parcels. Given the vestibular node's expected functional specificity, this separation appears more consistent with known vestibular network organization (Frank and Greenlee, 2018) and suggests that SWpC may better preserve functionally distinct pathways that are blurred in purely correlational representations.

Second, group differences depended on the feature type. Fractional occupancy did not differ significantly between HC and subacute PCVD for strength-based states, whereas duration-based measures showed a significant difference in one state. However, when assessing discriminability at the subject level, the strongest HC–ST separation emerged from strength-based features: state 5 was repeatedly selected as most discriminative for both SWC- and SWpC-strength, but SWpC-strength achieved substantially higher AUC than SWC-strength. In contrast, duration-based classification performance was near chance, which may reflect the greater variability of window-wise duration estimates and limited statistical power in this modest sample (n = 24 per group). Taken together, these results suggest that, in this dataset, directed strength carries the dominant discriminative signal, while duration may capture complementary group effects (e.g., occupancy shifts) without providing robust separability for classification. This pattern mirrors the motor-task results, where SWpC-strength also showed greater sensitivity than SWC-strength for detecting task-evoked changes, supporting the broader conclusion that prediction-based directed strength is a particularly informative and stable marker across contexts.

### 3.4. Methodological considerations and limitations

Several limitations should be considered when interpreting these findings. First, SWpC embeds a relatively simple linear time-invariant (LTI) model within each window. While this choice supports interpretability and computational tractability, it is ultimately a proxy for directed interactions that may be nonstationary and/or nonlinear, with time-varying latency, state-dependent gain, or interaction shapes that are not well captured by a window-wise LTI approximation. Relatedly, enforcing a fixed window length and overlap can constrain how rapidly the model adapts when interaction strength and temporal profile change at different rates across the scan. A promising direction is to incorporate adaptive window length so that the analysis can allocate shorter windows to rapidly changing regimes and longer windows to more stable periods.

In parallel, extending SWpC to nonlinear time-varying causal system models (e.g., locally nonlinear predictors, kernelized or state-space formulations) could better capture elastic, context-dependent directed interactions while retaining the windowed workflow. Finally, the PCVD dataset is relatively small; although nested cross-validation mitigates information leakage, effect sizes and discriminability should be re-evaluated in larger cohorts.

## 3.5. Outlook

Overall, SWpC provides a practical expansion of the widely used sliding-window workflow by adding directionality and temporal extent of influence. Across multimodal validation and human task and patient applications, SWpC yields stable estimates, improves sensitivity to condition-dependent changes, and offers an interpretable decomposition of dynamic network organization. The observed dissociation between strength and duration further motivates treating these features as complementary axes of directed connectivity, which may prove useful for probing mechanistic hypotheses and for developing clinically relevant biomarkers as datasets grow at scale.

# 4. Methods

## 4.1. Simultaneously recorded LFP-BOLD data in lightly anesthetized rats

This study analyzed data from 22 simultaneous recordings of single-slice functional magnetic resonance imaging (fMRI; TR = 0.5 s) and bilateral primary somatosensory cortex (S1) local field potentials (LFPs) from 10 male Sprague–Dawley rats (200–300 g; Charles River Laboratories) under dexmedetomidine (DMED) anesthesia, a regime that approximates a lightly sedated or near-awake state (REF). This dataset was originally acquired in multiple previous studies (Pan et al., 2011b) and processed following procedures detailed in (Zhang et al., 2020), which are summarized below.

All experiments were approved by the Emory University Institutional Animal Care and Use Committee and adhered to NIH guidelines. Under DMED anesthesia, fine-tip electrodes (~10 μm diameter; impedance 1–5 MΩ) were bilaterally implanted in the forelimb regions of S1. Simultaneous single-slice fMRI (9.4 T Bruker MRI system) and bilateral S1 LFP recordings were acquired. Functional imaging targeted a coronal slice covering bilateral S1 regions, with parameters: field of view (FOV) = 1.92 × 1.92 cm², matrix size = 64 × 64, in-plane resolution = 0.3 × 0.3 mm², slice thickness = 2 mm, echo time (TE) = 15 ms, and repetition time (TR) = 500 ms. Each scan lasted 8 min 20 s (1,000 TRs) plus 20 dummy scans for stabilization.

LFP data were low-pass filtered at 100 Hz, downsampled to 500 Hz, and synchronized to fMRI volumes. Gradient artifacts were removed using a template-based subtraction method (Pan et al., 2011b). The cleaned LFPs were band-pass filtered between 0.1 and 100 Hz and divided into six frequency bands: δ (1–4 Hz), θ (4–8 Hz), α (8–12 Hz), low β (12–25 Hz), high β (25–40 Hz), and γ (40–100 Hz). Band-limited power (BLP) time courses were computed using Welch's method with a 1 s (2 TRs) sliding window and 50% overlap, then normalized for comparability.

fMRI data underwent standard preprocessing, including motion correction and Gaussian spatial smoothing (FWHM = 0.84 mm). Global signals and linear drifts were regressed out, and temporal band-pass filtering (0.01–0.25 Hz) was applied to reduce noise. Subsequently, 22 selected scans were normalized and ROI-based BOLD signals were z-scored. To align with the hemodynamic response function peak latency under DMED (~2.5 s), BOLD signals were shifted by 2.5 s relative to LFP-derived BLP signals, resulting in 995 TRs (BOLD: [6:1000], BLP: [1:995]) for direct comparison across all frequency bands.

## 4.2. Human motor task fMRI data from the Human Connectome Project (HCP)

Motion task-evoked fMRI (tfMRI) data from the Human Connectome Project (HCP) test-retest cohort were used to assess SWpC sensitivity. The chosen motor task induces robust activations in motor and somatosensory cortices as well as the cerebellum (Barch et al., 2013). Minimally preprocessed 3T motor task fMRI data from the test-retest group was obtained from the Human Connectome Project (HCP). The study included 45 test-retest subjects, each completing two motor task runs. During each run, participants performed five visually cued motions—left foot, right foot, left hand, right hand, and tongue—each repeated twice, resulting in four trials per motion across both runs. Each trial lasted 12 seconds and was preceded by a 3-second visual cue. Additionally, two 12-second resting trials, occurring after one repetition of the five motions, were designated as the null condition for subsequent analysis (Fig. S1).

Task-specific activation maps for the motor task were derived using a three-level FEAT analysis. At *Level 1*, activation maps within each run were identified using a General Linear Model (GLM) as described in (Barch et al., 2013). Five movement predictors—left foot (lf), right foot (rf), left hand (lh), right hand (rh), and tongue (t)—were convolved with a double gamma hemodynamic response function (Glover, 1999), with temporal derivatives included as confounds. The minimally preprocessed fMRI data further underwent spatial smoothing with a FWHM of 4 mm, high-pass filtering with a 200 s cutoff, and prewhitening using FILM to correct for autocorrelations (Woolrich et al., 2001). At *Level 2*, both runs of the Level 1 results for each subject were combined using fixed-effects modeling to produce a single activation map per subject; cluster thresholding was applied with a Z threshold of ±3.29 and a cluster p-threshold of 0.05. At *Level 3*, data across subjects were aggregated using a random-effects model to create group-level activation maps for each task, with cluster thresholding applied using a Z threshold of ±4.42 and a cluster p-threshold of 0.001.

For all three-level analyses, the GLM design included task contrasts for visual cue, each motor task, and each motor task versus baseline. Baseline was defined as the average of all other task conditions (e.g., left hand vs. the average of right hand, left/right foot, and tongue), a method shown to optimize task specificity in brain activation maps (Tripathi et al., 2024). Group-level activation maps for each task contrast relative to its baseline demonstrated consistent task-related activations in the brain. These regions are referred to as task-specific regions of interest (ROIs) throughout this paper. Time series were extracted from these task-specific ROIs for each motion and resting trial, yielding four task time series and four resting time series per subject.

### 4.3. Resting fMRI data in PCVD patients

Post-concussion vestibular dysfunction (PCVD) was hypothesized to involve disrupted directed functional connectivity (FC) along visual–vestibular multisensory processing pathways (Smith et al., 2021). Subacute PCVD (ST) was defined as occurring 2–12 weeks post-concussion and is often considered the stage most responsive to vestibular rehabilitation, whereas chronic PCVD (CH) was defined as vestibular symptoms persisting for >6 months after the concussion. In contrast, healthy controls (HC) had no history of concussion-related vestibular dysfunction or other vestibular impairment. The primary analyses in this study focus on 24 HC versus 24 ST participants. An auxiliary chronic PCVD (CH) cohort (n=24) was included solely to improve the stability of unsupervised brain-state estimation used for state-based feature extraction. All primary hypothesis testing and predictive analyses reported here were restricted to HC vs subacute PCVD (see Section Methods—Functional demonstration on PCVD patients and clinical prediction for details).

Inclusion criteria for PCVD patients required a diagnosis of concussion as defined by the World Health Organization Collaborating Center for Neurotrauma Task Force and clinical evidence of vestibular impairment. Vestibular symptoms were confirmed through subjective reports of dizziness or imbalance, visual motion sensitivity, and symptom provocation during Vestibular/Ocular-Motor Screening (VOMS). Exclusion criteria for both healthy control and PCVD groups included a history of moderate or severe head injury, intracranial hemorrhage, seizure disorder, prior neurologic surgery, peripheral neuropathy, musculoskeletal injuries affecting gait or balance, or chronic drug or alcohol use. Participants with abnormal findings on head impulse testing or videonystagmography (VNG) indicative of peripheral vestibular hypofunction or benign paroxysmal positional vertigo were also excluded.

Resting-state fMRI (rs-fMRI) data were collected for all participants, with each scan lasting 420 seconds and using a repetition time (TR) of 0.7 seconds. Preprocessing was conducted using the CONN Toolbox v19c (Whitfield-Gabrieli and Nieto-Castanon, 2012) following the procedure outlined in (Smith et al., 2021). Standard steps included slice timing, field map correction, motion correction, and registration to the MNI152 template via DARTEL. Motion parameters, outliers, and mean CSF and white matter signals were regressed out. The data were then smoothed with an 8 mm FWHM Gaussian kernel and band-pass filtered (0.01–0.25Hz). All processed data and masks were manually inspected for accuracy, with maximal framewise displacement recorded as 0.76 mm. Timeseries were extracted from 26 parcels of EAGLE449 atlas that cover the core and extended vestibular networks (Smith et al., 2023, Table 1). The extracted timeseries were despiked and low-pass filtered with a cutoff frequency of 0.15 Hz. This dataset provided a foundation for assessing dynamic functional connectivity and spatiotemporal patterns of brain activity, particularly centered around the vestibular networks.

### 4.4. Reliability validation on concurrent LFP-BOLD data in rats

To validate the reliability of SWpC in estimating dynamic directional functional connectivity (FC), we analyzed 22 concurrent LFP-BOLD recordings from the somatosensory cortices (S1L and S1R) of rats (as described in Section 4.1). The physiological expectation of symmetrical information flow between S1L and S1R served as a benchmark for model validation. SWpC was applied to both BOLD signals and band-limited power (BLP) signals across six frequency bands ($\delta$, $\theta$, $\alpha$, low $\beta$, high $\beta$, and $\gamma$) to evaluate its ability to (1) capture time-varying directional FC at both BOLD and neuronal levels and (2) assess whether the strength and/or duration of information flow detected via BOLD signals was associated with neuronal activity in specific frequency bands. Results were contrasted with metrics derived from traditional sliding window correlation (SWC). Following (Garth John Thompson et al., 2013a), a 50-second sliding window that slides every time point was used for both SWpC and SWC analyses to ensure consistency and facilitate direct comparisons. Each window encompassed 100 timepoints, with a maximum duration of interest set at 15 seconds (or 30 timepoints). To select the optimal model configuration and minimize the risk of overfitting, the Bayesian Information Criterion (BIC) was employed in estimating SWpC durations. This approach enhanced the accuracy and reliability of the duration estimates.

To evaluate SWpC's performance in estimating time-varying directional FC (S1L→S1R vs. S1R→S1L), we quantified directional asymmetry by calculating the difference in SWpC strength and duration measures between the efferent (S1L→S1R) and afferent (S1R→S1L) directions. For each scan, we first computed the mean SWpC strength or duration across all sliding windows, then defined directional asymmetry as the difference between these two directions. Scan-to-scan variability was determined as the standard deviation of these mean values across all scans for each signal type, including BOLD and BLP signals across six frequency bands ($\delta$, $\theta$, $\alpha$, low $\beta$, high $\beta$, and $\gamma$). Strength represented the amplitude of directional interactions, while duration reflected how long these interactions persisted within each sliding window. The Wilcoxon signed-rank test was applied to determine whether the median directional asymmetry was significantly lower than the scan-to-scan variability. This non-parametric approach provided a robust assessment of the stability of SWpC-derived estimates, ensuring that results were not disproportionately influenced by deviations from normality or outliers.

We further investigate the neuronal contributions to BOLD-directed FC by correlating SWpC-derived BOLD metrics with BLP signals from six frequency bands. This analysis aimed to clarify whether directional connectivity observed in BOLD signals originated from specific neuronal frequency ranges. To contextualize these findings, we compared SWpC-based correlations with those derived from SWC, highlighting SWpC's unique capability in characterizing the neural underpinnings of directional information flow. Due to the large variations observed across different scans, a bootstrapping approach was implemented to account for this variability. In each bootstrap iteration, the 22 fMRI scans were randomly divided into two cohorts. For each cohort and each frequency band ($\delta$, $\theta$, $\alpha$, low $\beta$, high $\beta$, and $\gamma$), Fisher z-transformed correlations were computed between the concatenated sliding window timecourses of BOLD and band-limited power (BLP) signals. This procedure was repeated over 10,000 bootstrap iterations to generate distributions of correlation values for each frequency band. A box plot was employed to visualize the bootstrap distributions, with each box representing the interquartile range (IQR) of the correlations, whiskers extending to 1.5 times the IQR, and outliers displayed as individual points. This approach enabled a robust evaluation of correlation variability while incorporating random variations introduced by cohort splitting.

### 4.5. Sensitivity validation on motor HCP data

To assess the sensitivity of SWpC, we applied it to motor task data from the Human Connectome Project (HCP), which elicits strong directed functional connectivity (FC) between task-specific regions of interest (ROIs). SWpC results were compared with those obtained using standard sliding window correlation (SWC). Each 12-second task or resting trial was treated as an individual window, and both methods were applied to the motion and resting time series for all subjects and runs. Given the small sample size per window, we utilized the corrected Akaike Information Criterion (AICc) for model selection, as it outperforms AIC and BIC under these conditions (Hurvich and Tsai, 1989). For each subject, SWC generated four time-varying strength matrices, and SWpC produced four strength and four duration matrices for each motion-task and corresponding resting condition. Averaging across the four trials of each condition resulted in one time-averaged SWC strength matrix and one SWpC strength and duration matrix per condition.

To assess the asymmetry between directed functional connectivity (FC) results for rest versus task conditions, we quantified asymmetry for each subject's time-averaged matrix by calculating the Frobenius norm of the difference between the matrix and its transpose, which reflects the overall deviation from symmetry. We conducted normality tests on these asymmetry measures across all subjects to determine the appropriate statistical tests—either independent two-sample t-tests or Mann-Whitney U tests based on data distribution—and calculated Cohen's d to quantify the magnitude of the observed differences.

Directed FC changes between task and rest conditions were examined for both duration and strength measures. For time-averaged directed FC duration matrices estimated by SWpC, we identified significant differences in directed FC durations between task and rest conditions. Statistical tests were performed on each matrix entry to determine whether these differences were significant (FDR-corrected p-value < 0.01) and exhibited low individual variability (coefficient of variation [CV] < 36%). This approach pinpointed specific connections with significant task-driven changes in the duration of directed FC.

To examine differences in the strength of directed FC between task and rest conditions, we compared time-averaged SWpC results with SWC results by analyzing two pairs of matrices: (directed FC_task, directed FC_rest) and (standard FC_task, standard FC_rest). Element-wise differences for each pair, defined as (directed FC_task - directed FC_rest) and (standard FC_task - standard FC_rest), were computed across all subjects. Statistical approaches were used to quantify these differences. First, significance tests were conducted to identify task-specific connections with low individual variability. For each entry in the difference matrices, a t-test determined whether the mean difference significantly deviated from zero (p < 0.01, FDR-corrected). Among significant connections, those with low individual variability (CV < 30%) were selected. Second, we evaluated the relative sensitivity of SWpC and SWC strengths in detecting task-specific connections. Specifically, the Frobenius norm, defined as the square root of the sum of squared elements in each difference matrix, was calculated to provide a scalar measure of overall distinctions between task and rest. The mean Frobenius norm across all subjects for each matrix pair represented the difference on average. A two-sample t-test was then conducted to determine whether one pair showed significantly larger differences than the other. Additionally, the relative difference between pairs was expressed as the percentage difference in their mean Frobenius norms. This analysis facilitated a direct comparison of SWpC and SWC in detecting task-specific changes in directed FC strength.

Furthermore, we assessed whether time-varying directed FC strength captured trial-to-trial dynamics across the four repetitions of each motion task. We quantified agreement between methods by correlating SWpC- and SWC-derived strength estimates across trials (visualized with a scatter plot). We also examined behavioral relevance by correlating, across the four trials, group-averaged mean directed FC strength (average connectivity across all ROI pairs) with group-averaged reaction time (RT) using Pearson correlation, noting that inference is limited by the small number of time points (n = 4).

### 4.6. Functional demonstration on PCVD patients and clinical prediction.

The time-varying causal interactions estimated by SWpC were demonstrated on the rsfMRI data of PCVD patients which are compared with the SWC results. Following previous human study (Allen et al., 2012), we applied a 44-s tapered sliding window that advanced by one TR at each step, generating a time-resolved sequence of SWC and SWpC connectivity matrices for every subject. To obtain a stable, common set of recurring dynamic patterns, the resulting matrices from all available subjects (HC, ST, and CH) were concatenated and submitted to k-means clustering to derive shared brain-state centroids. The Elbow criterion indicated that five clusters provided the most stable decomposition; thus, the cohort was characterized using a common set of five dynamic brain states.

All subsequent inferential and predictive analyses reported in this study were restricted to the primary comparison of HC versus subacute PCVD (ST). After the shared state centroids were estimated, each subject's windows were assigned to the five states, and we computed (i) connectivity strength, defined as the average edge weight across all ROI pairs within windows assigned to a given state, and (ii) fractional occupation (FO), defined as the proportion of windows belonging to each state. These features captured both the spatial profile and temporal expression of the dynamic patterns and were used for downstream analyses focusing on HC and ST.

To further investigate the large-scale network structure embedded within these dynamic patterns, we averaged the SWC and SWpC matrices across time and applied the Louvain community detection algorithm to the resulting functional networks. Group-level comparisons of modular organization and network segregation/integration were performed for HC versus ST only, in order to assess whether SWpC and SWC highlighted different vestibular–multisensory pathway organization and whether subacute PCVD showed altered network topology relative to controls.

Finally, to evaluate the clinical relevance of state-derived features given the limited patient dataset, we implemented a nested leave-one-out cross-validation (LOOCV) machine-learning framework (Vergara et al., 2018) to discriminate subacute PCVD (ST) from HC. In each fold, one participant was held out as the test case, and a support vector machine (SVM) classifier was trained on the remaining participants using the features derived from the five states for each of the three cases: SWC strength, SWpC strength, and SWpC duration. Within each training fold, features were ranked by their two-sample t-test statistic between groups, and the top [2%: 2%: 20%] were retained. The optimal threshold was selected within the inner loop of the nested CV procedure. The resulting classifier was evaluated on the held-out participant, and performance was quantified using the area under the ROC curve (AUC). This nested design prevented information leakage between training and testing and provided an unbiased estimate of the ability of dynamic connectivity features to discriminate subacute PCVD patients from healthy controls.

## 5. Data and code availability statement

Our code for computing SWpC is available at https://github.com/inspirelab-site/swpc.

## 6. Credit authorship contribution statement

**Nan Xu:** Conceptualization, Data curation, Data preprocessing, Methodology, Formal analysis, Writing – original draft, Writing – review & editing, Funding acquisition. **Xiaodi Zhang:** Data curation, Data preprocessing for rodent data, **Wen-Ju Pan:** Rodent data acquisition. **Jeremy L. Smith:** Data curation, Data preprocessing for PCVD data. **Eric H. Schumacher:** Writing – review & editing. **Jason W. Allen:** Writing – review & editing, Supervision. **Vince D. Calhoun:** Writing – review & editing, Supervision. **Shella D. Keilholz:** Writing – review & editing, Supervision.

## 7. Declaration of Competing Interest

The authors declare that they have no situation of real, potential or apparent conflict of interest and that there is no financial/personal interest or belief that could affect their objectivity.

## 8. Acknowledgement

Nan Xu thanks the funding support for National Institutes of Health (NIH K99/R00NS123113). Nan Xu and Shella D. Keilholz also thank the funding support from National Institutes of Health (NIH R01NS078095).

Supplementary Material

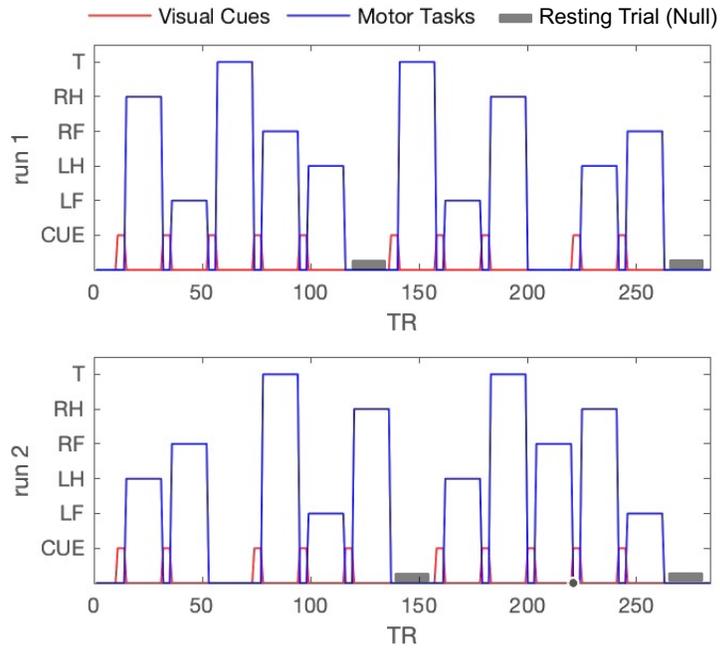

Fig. S1. Timing of the visually cued motion tasks and corresponding resting trials (null conditions) for each fMRI run.

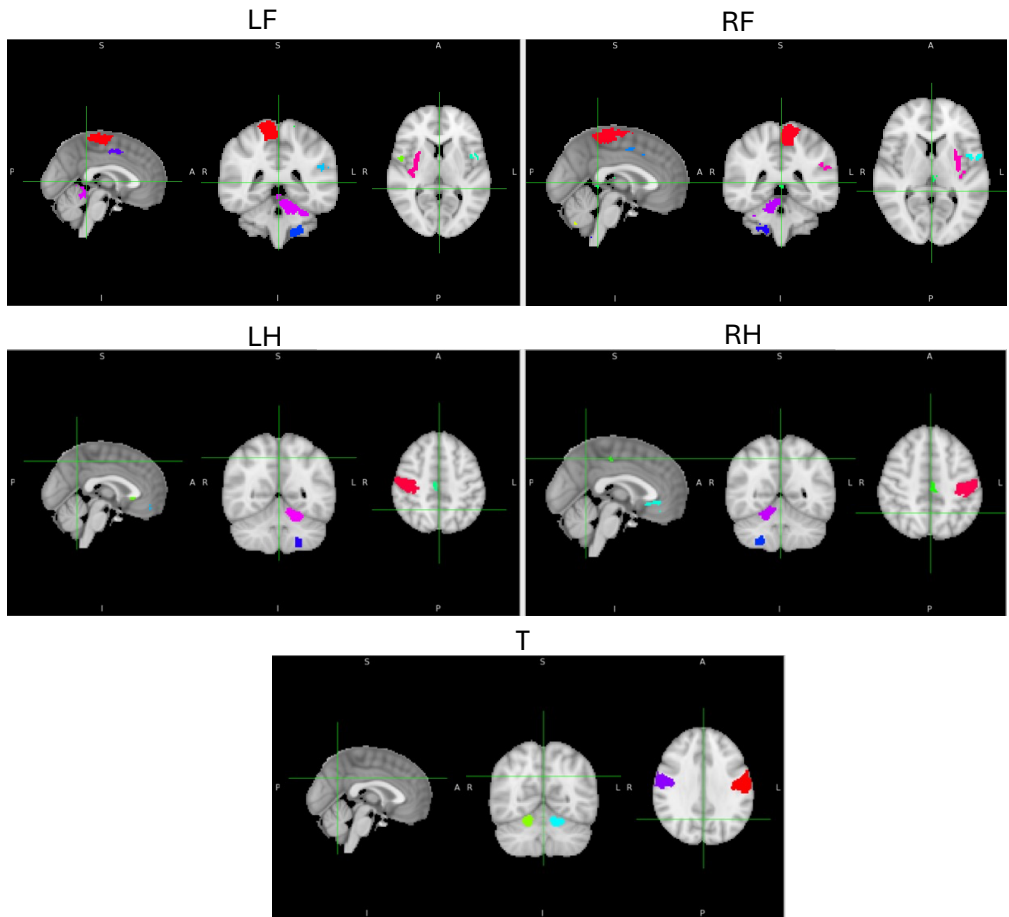

Fig. S2. Activation maps (ROIs) derived from FEAT analysis for each motion task.
0

Supplementary Material

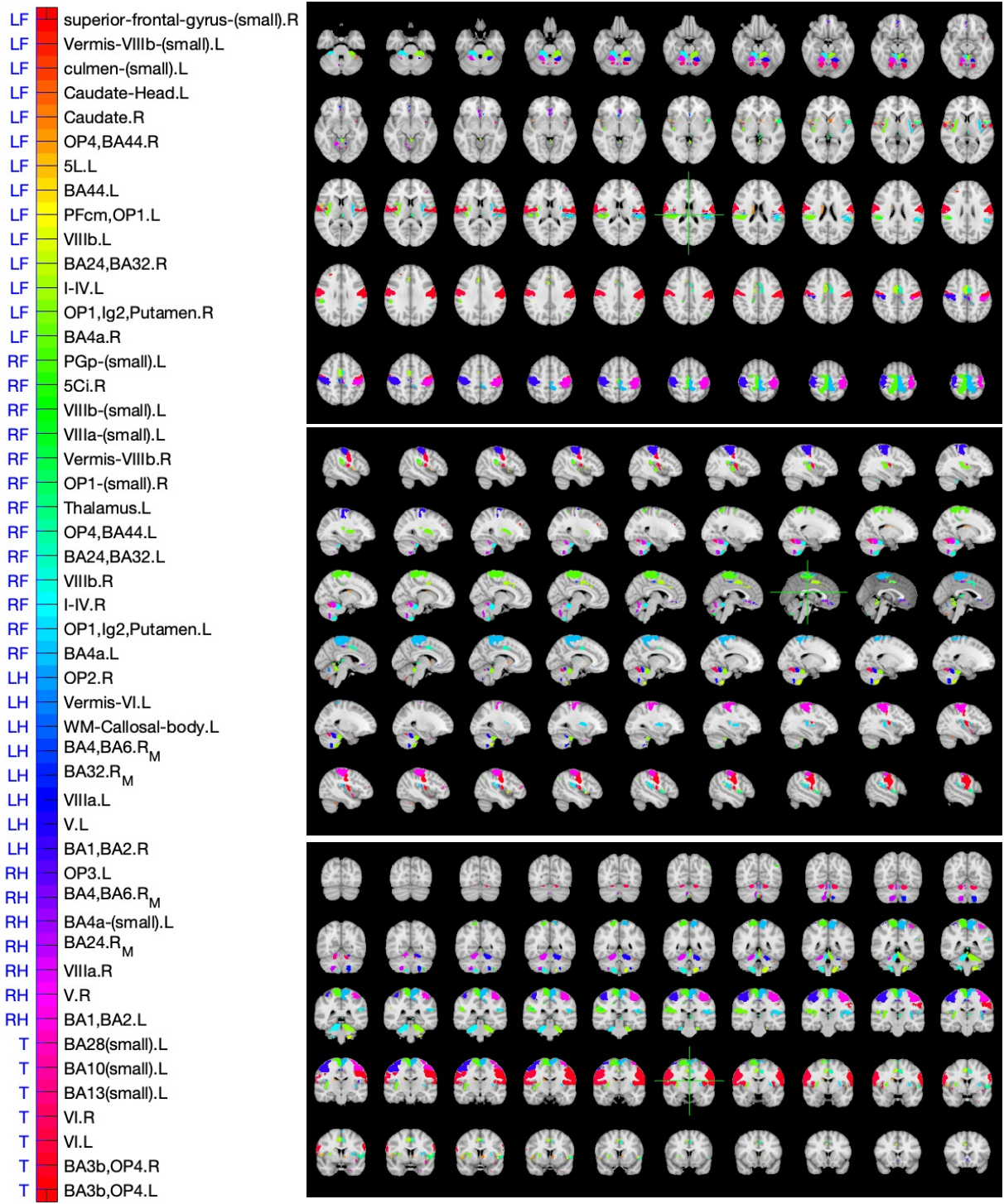

Fig. S3. 49 motion-specific ROIs identified by 3-level FEAT analysis

Supplementary Material

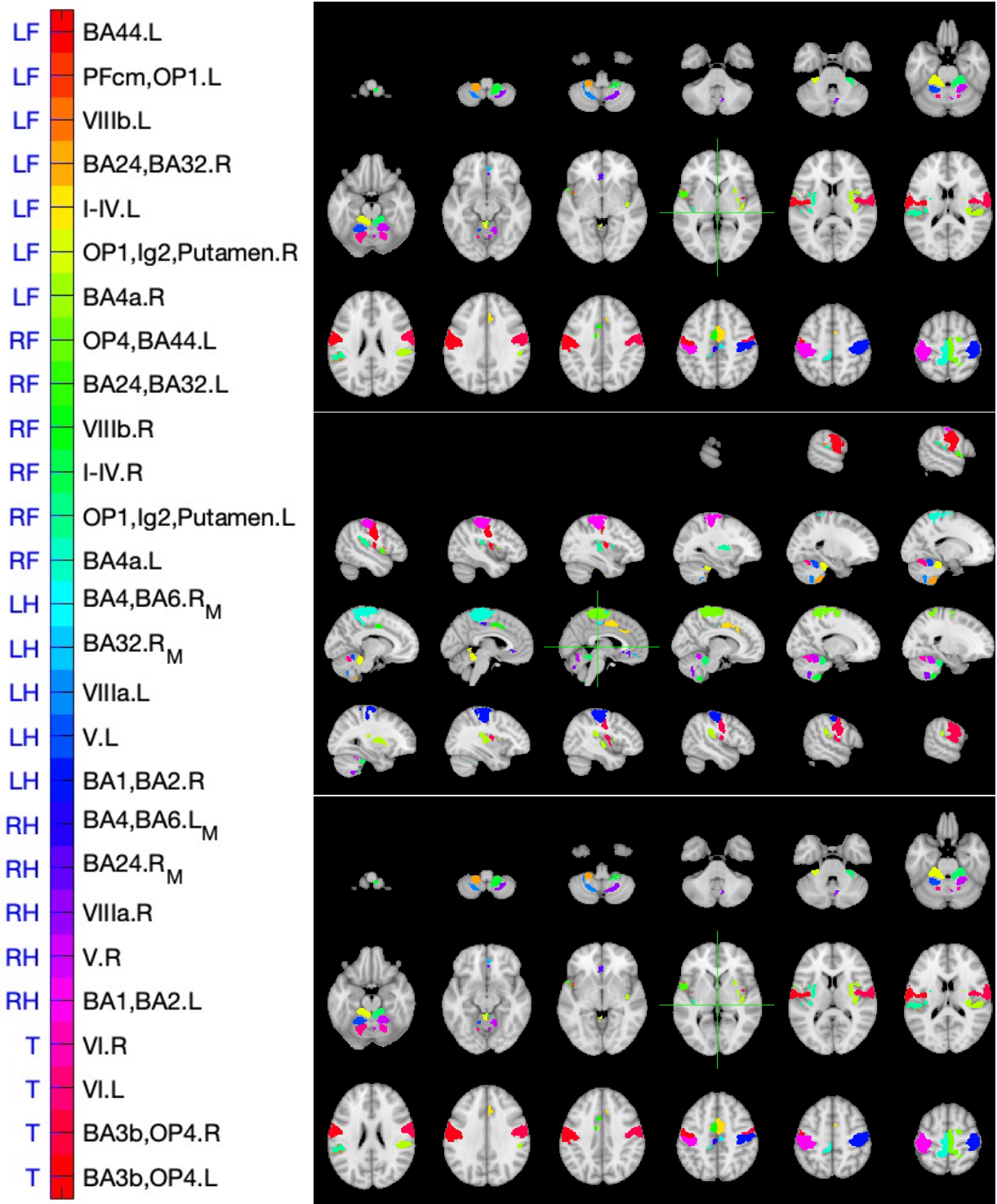

Fig. S4. 27 selected large motion-specific ROIs for illustration purposes. These include 13 foot ROIs, 10 hand ROIs, and 4 tongue ROIs, with smaller ROIs excluded (fewer than 100 voxels for foot motions and fewer than 30 voxels for hand and tongue motions). These 27 ROIs exhibit approximate symmetry for left/right foot and hand tasks.

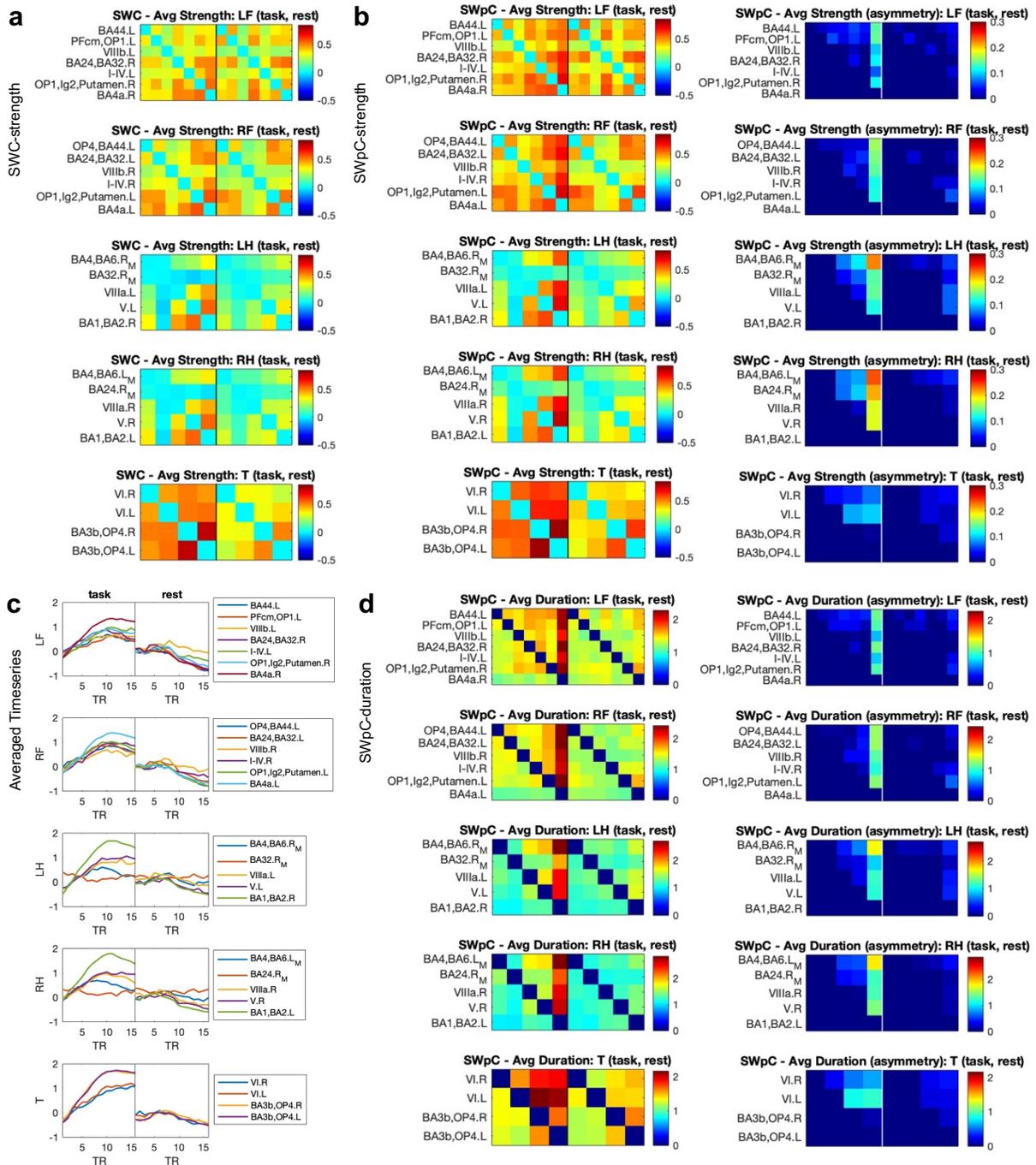

Fig. S5. Group-averaged (directed) FC strength and duration matrices as well as fMRI timeseries during task and rest trials.

Supplementary Material

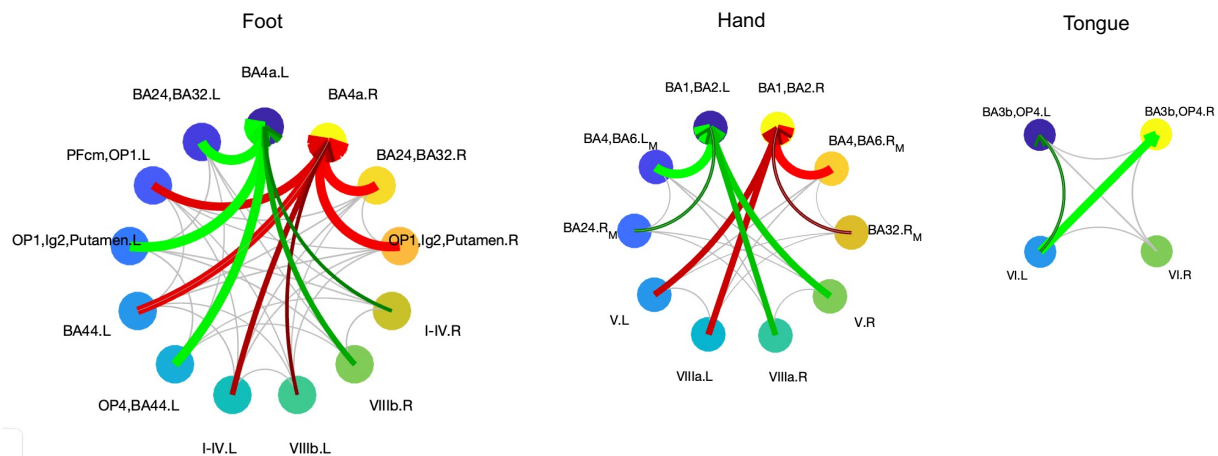

Fig. S6. Connectograms of directed FC with significant differences in duration between task and rest for hand (left), (middle), and tongue (right) motions. Red (green) arrows indicate task vs. rest differences for left (right) motion with low individual variability (CV < 36%). For tongue motions, green arrows represent significant evoked directed FC strengths with low variability. Gray connections denote all predicted directed connections from SWpC. To enhance clarity, connectograms include 13 foot ROIs, 10 hand ROIs, and 4 tongue ROIs, excluding smaller ROIs (fewer than 100 voxels for foot motions and fewer than 30 voxels for hand and tongue motions). The remaining ROIs exhibit approximate symmetry for left/right foot and hand tasks. Under these thresholds, tongue-evoked information flow among the 4 larger ROIs does not form hubs in the motor cortex but connects VI.L to two somatomotor cortical regions, potentially due to the applied threshold amount the limited number of connections.

|  | Motion | Distribution Type | Asymmetry Measure (Task) | Asymmetry Measure (Rest) | P-value | Cohen's D effect size |
|---|---|---|---|---|---|---|
| strength | LF | Normal | 1.87 | 1.50 | 1.53E-06 | 1.0712 |
|  | RF | Normal | 1.66 | 1.39 | 1.87E-06 | 1.0499 |
|  | LH | Normal | 1.13 | 0.87 | 3.23E-06 | 1.0223 |
|  | RH | Normal | 1.20 | 0.71 | 2.43E-12 | 1.7892 |
|  | T | Normal | 0.99 | 0.68 | 8.03E-07 | 1.1007 |
| duration | LF | Normal | 10.21 | 7.31 | 8.19E-08 | 1.246 |
|  | RF | Normal | 9.45 | 6.49 | 3.06E-13 | 1.8447 |
|  | LH | Normal | 6.66 | 3.91 | 8.99E-13 | 1.8305 |
|  | RH | Normal | 7.33 | 3.14 | 3.15E-15 | 2.1781 |
|  | T | Normal | 6.67 | 3.32 | 1.37E-13 | 1.9559 |

Table S1. Directional Asymmetry Measures in SWpC strength and Durations.

Supplementary Material

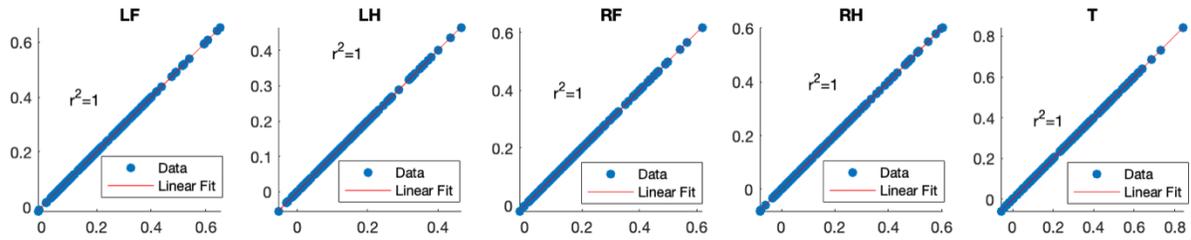

Fig S7. Scatter Plot of SWPC-Directed FC Matrix Strength vs. SWC FC Matrix Strength

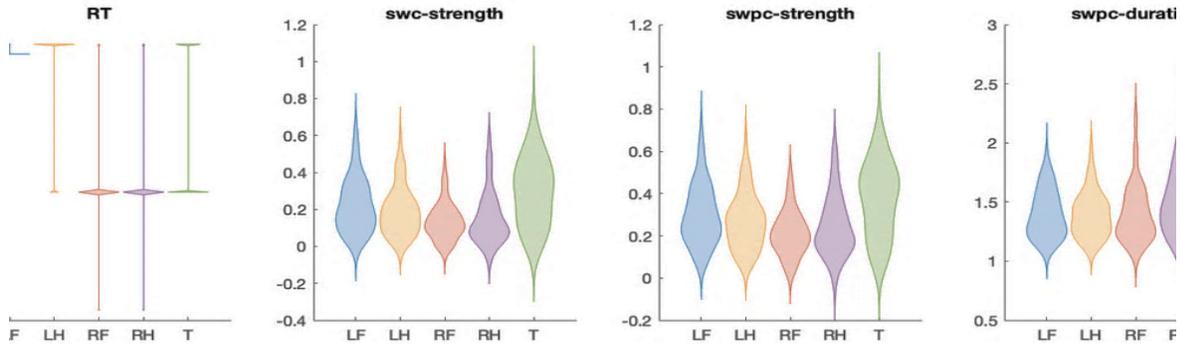

Fig S8. Violin plots of reaction time (RT), and SWC as well as SWpC estimates.

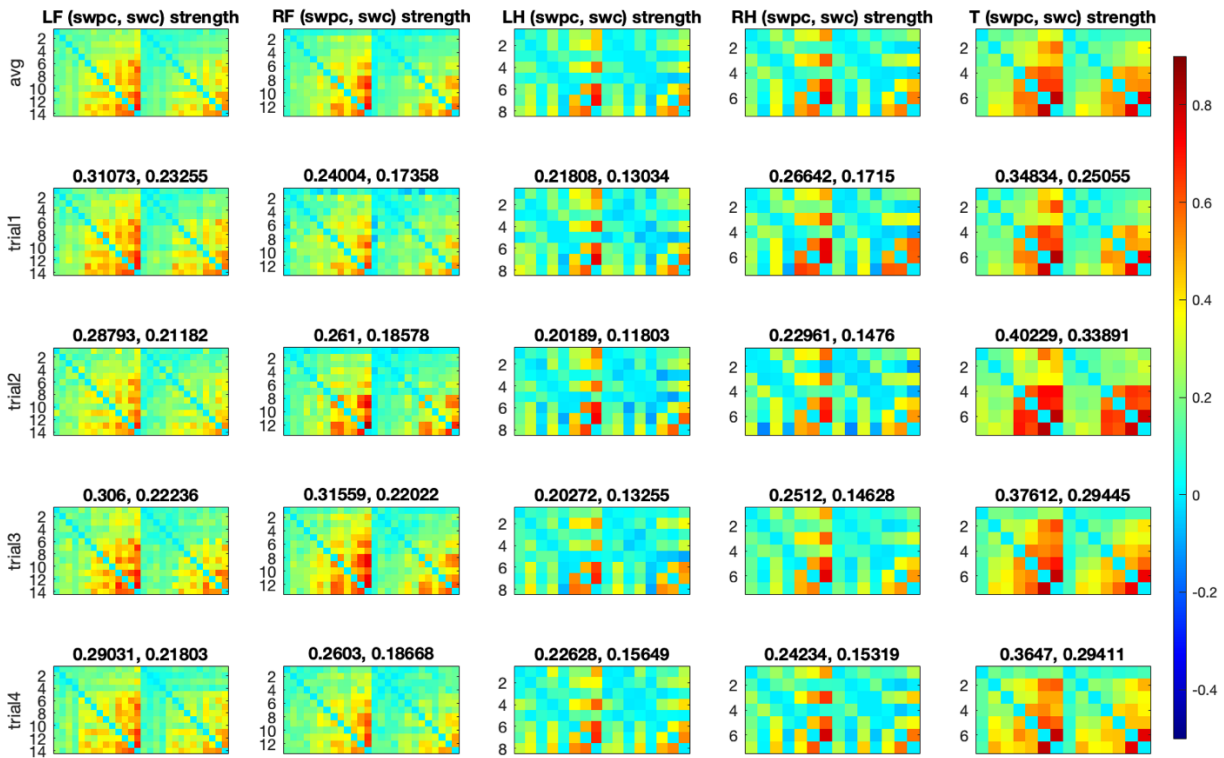

Fig S9. Group-averaged SWpC and SWC strength during each task trial. The mean FC strength was listed on the top of each matrix.
5

Supplementary Material

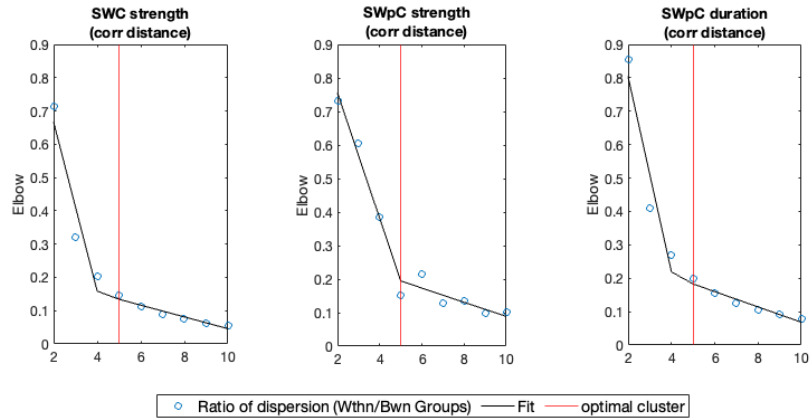

Fig S10. Optimal k-means cluster number by Elbow method.

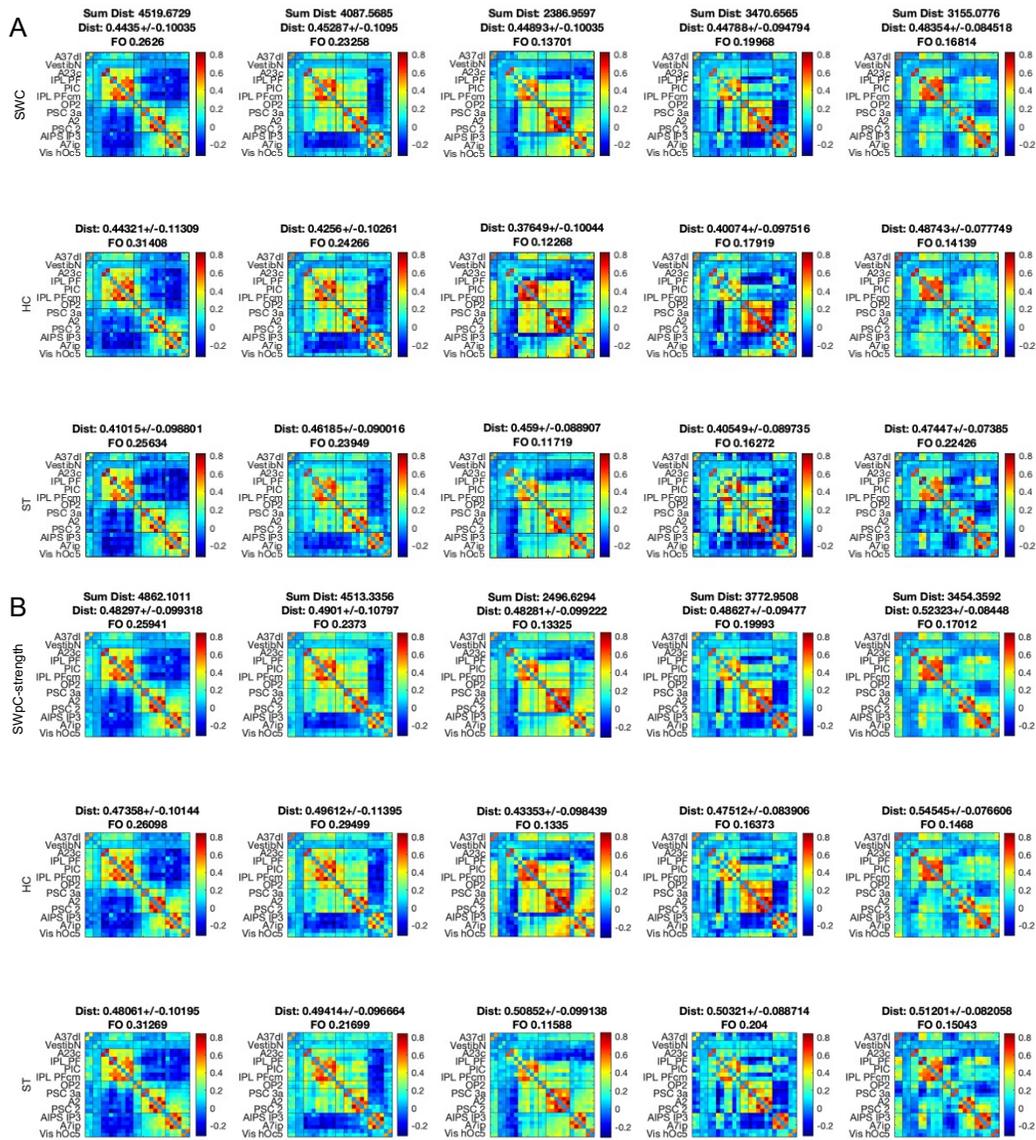

Supplementary Material

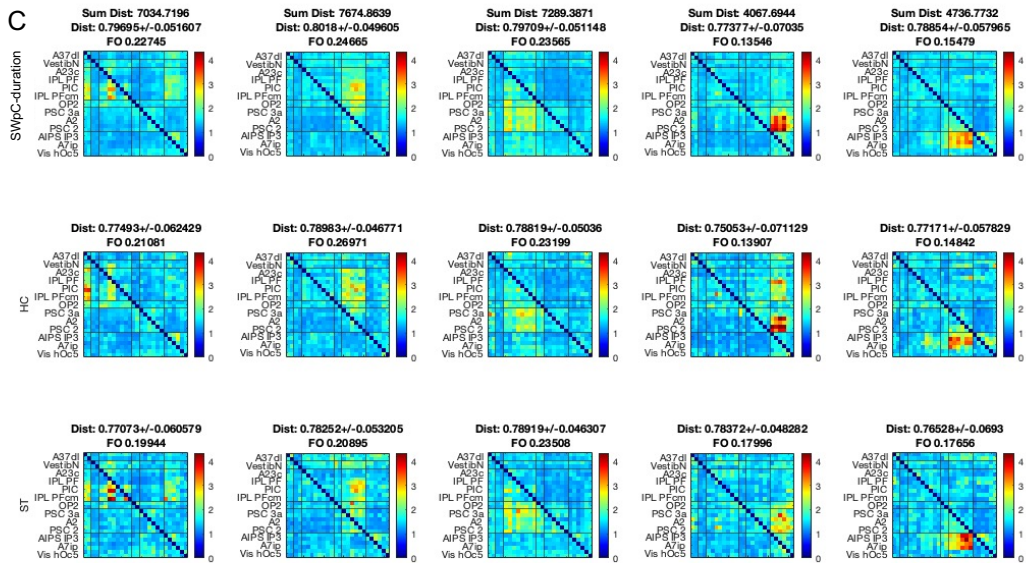

Fig. S12. Brain states estimated using (A) SWC, (B) SWpC-strength, and (C) SWpC-duration, shown for states shared across HC, ST, and CH (top), as well as states specific to HC (middle) and ST (bottom).

|    | State 1   | State 2   | State 3   | State 4   | State 5   |
|----|-----------|-----------|-----------|-----------|-----------|
| HC | 27% (23)  | 29% (24)  | 14% (17)  | 16% (16)  | 14% (21)  |
|    | 26% (23)  | 30% (23)  | 13% (20)  | 16% (17)  | 15% (20)  |
|    | 22% (24)  | 30% (24)  | 22% (24)  | 13% (24)  | 14% (24)  |
| ST | 31% (23)  | 21% (22)  | 11% (19)  | 21% (22)  | 15% (20)  |
|    | 31% (23)  | 22% (22)  | 12% (19)  | 20% (21)  | 15% (20)  |
|    | 23% (24)  | 21% (24)  | 24% (24)  | 13% (24)  | 19% (24)  |

Table S2 Fraction of occupation for each state for each group with parathesis including the number of subjects that visit each state in each group. Color-code (black, blue and purple) enlists the results of swc-strength, swpc-strength, swpc-duration.

Supplementary Material

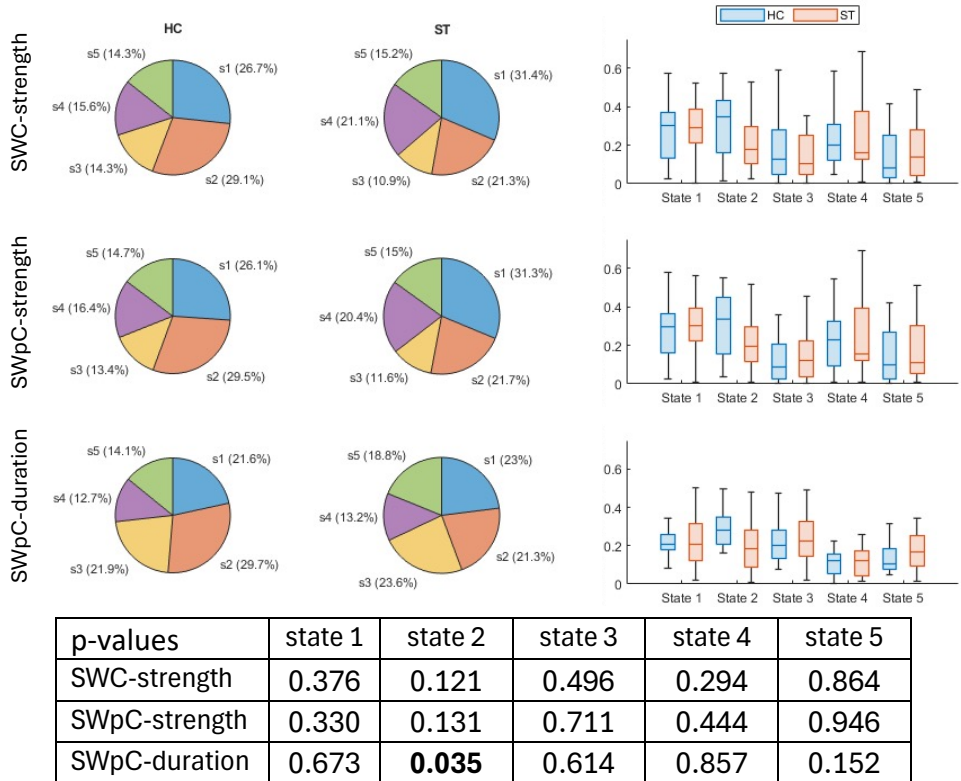

| p-values | state 1 | state 2 | state 3 | state 4 | state 5 |
|---|---|---|---|---|---|
| SWC-strength | 0.376 | 0.121 | 0.496 | 0.294 | 0.864 |
| SWpC-strength | 0.330 | 0.131 | 0.711 | 0.444 | 0.946 |
| SWpC-duration | 0.673 | **0.035** | 0.614 | 0.857 | 0.152 |

Fig. S13. Fractional occupancy in Healthy Controls and subacute PCVD patients. Left: pie charts based on group mean values. Right: bar plots showing the group means and variability.